\documentclass[11pt,a4paper]{article}
\usepackage{authblk}
\usepackage[english]{babel}
\usepackage{amssymb,amsmath}

\newcommand{\be}{\begin{equation}}
\newcommand{\ee}{\end{equation}}
\newcommand{\bea}{\begin{eqnarray}}
\newcommand{\eea}{\end{eqnarray}}

\def\a{\alpha}

\def\r{\rho}

\def\br{\bar{\rho}}
\def\bg{\bar{g}}
\def\bpi{\bar{\pi}}

\def\bra{\langle}
\def\ket{\rangle}

\def\sst{\scriptscriptstyle}

\begin{document}

\title{{\bf Multiparticle correlation expansion of relative entropy in lattice systems}
\thanks{J. Stat. Mech. 073201 (2016) doi:10.1088/1742-5468/2016/07/073201}}
\author{Marco D'Alessandro\footnote{e-mail address: marco.dalessandro@isc.cnr.it}}
\affil{\emph{Institute for Complex Systems, National Research Council (CNR), Via del
Fosso del Cavaliere 100, 00133 Rome, Italy}}
\date{}

\maketitle
\vspace{-1cm}
\begin{abstract}
This paper deals with the construction of the multiparticle correlation expansion of relative entropy for lattice systems. Thanks to this
analysis we are able to express the statistical distance between two systems as a series built over clusters of increasing dimension. Each
addend is written in terms of correlation functions and expresses the contribution to the relative entropy due to structural information inside
the selected cluster. We present a general procedure for the explicit construction of all the terms of the series. As a first application of this
result, we show that the coefficients of the multiparticle correlation expansion of the excess entropy can be computed from our formula, as a
particular case.
\end{abstract}

%%%%%%%%%%%%%%%%%%%%%%%%%%%%%%%%%%%%%%%%%%%%%%%%%%%%%%%%%%%%%%%%%%%%%%%%%%%%

\section{Introduction}
\label{Intro}

Multiparticle correlation expansion in the framework of equilibrium statistical mechanics is a useful tool since it allows to predict
thermodynamic properties from structural correlation functions. The aim of the multiparticle correlation expansion is to express the
thermodynamic potentials as a series made of integrals defined over subset of variables (cluster). The density of the system weights the
elements of the series associated to clusters of increasing dimension and controls its convergence properties.
Classical examples of cluster expansion are given by the Mayer series for the partition function of a simple fluids, or by the correlation
expansion of the excess entropy in the continuum (see \cite{Puoskari:1999:0378-4371:509,Hernando1990,Prestipino2004}, and ref. therein,
for a modern derivation in both the canonical and grand-canonical ensemble) and on the lattice \cite{springerlink:10.1023/A:1004520432275}. These
papers have unveiled the interplay between entropy and spatial ordering evidencing that, in typical situations, two-body term provides a semi
quantitative estimate of the total entropy. Nonetheless, the residual entropy due to higher order correlations represents a rich source of information
concerning the thermodynamics of the system, in particular near phase transitions \cite{RePEc:eee:phsmap:v:187:y:1992:i:1:p:145-158,PhysRevA.45.R6966}.

Apart from these deeply investigated examples, there is an ulterior case that should deserve further study in this direction, namely the
analysis of the multiparticle correlation expansion of the relative entropy (RE) (also known as Kullback-Leibler divergence \cite{Kullback51klDivergence}
in the mathematical literature). RE has been firstly introduced in information theory as a notion of distance to quantify the dissimilarity
between two statistical systems \cite{Kullback51klDivergence}. Later on, inference methods based on the usage of RE have been developed
\cite{Heinz2005,Bavaud2009}. These procedures constitute a generalization the Maximum Entropy Principle of Jaynes \cite{Jaynes1957},
and aim to update a given ``prior'' distribution when new information, given in the form of some constraints, becomes available \cite{Janik:2003hk}.
Apart from these information theory oriented applications, RE possesses a nice physical meaning also in the thermostatistical framework as
it can be identified with the free energy difference between the thermodynamical equilibrium and the out of equilibrium fluctuations
\cite{PhysRevE.63.042103,PhysRevE.81.051133}.

Despite its potential applications in many different fields, the direct evaluation of RE turns out to be not achievable in many realistic
cases since it involves a sum over the entire state space of the system which is not computationally feasible when the number of degrees
of freedom is quite large. Cluster expansion reveals a useful tool to overcome this difficulty allowing to provide an estimate of RE built
over a lower dimensional system made of a selected group of clusters. Applications of this type of approach range from the cluster variation
methods for the evaluation of approximated free energy in lattice systems \cite{Guozhong1988,Pelizzola2005} to the usage of RE as a model
selection tool \cite{PhysRevB.87.174112}.

The aim of this paper is to derive the formal expression of the multiparticle correlation expansion of the relative entropy for systems on the lattice.
The considerations presented in the previous paragraphs highlight some of the reasons that motivate this kind of investigation. The choice of working
on the lattice is due to the fact that these systems costitute a useful framework for describing a wide range of physical models, from the Ising-like
models of statistical mechanics to the neural networks. Moreover, the corresponding results for systems in the continuum can be obtained by performing
the proper limit of the lattice variables.

The paper is organized as follows. Section \ref{PreliminaryResults} collects the definition of the relevant quantities together with some
background material. The derivation of the multiparticle correlation expansion of the relative entropy is discussed in section \ref{CERelativeEntropy}.
As a first application of this result we show that the coefficients of the entropy cluster expansion derived in \cite{springerlink:10.1023/A:1004520432275}
can be reproduced from our main result by performing a specific choice of the probability distributions. This is done in section \ref{EntropyCE}.
Finally, in section \ref{Conclusions} we discuss some concluding remarks.

\section{Preliminaries}
\label{PreliminaryResults}

\subsection{Basic notations and definitions}
\label{BasicNotations}

We consider a set of $N$ discrete variables $c_{i}$, with the index $i$ in the range $[1,N]$. The variables can assume the values
0 and 1. We indicate as a \emph{configuration} each specific assignment of the values of the $N$ variables, represented the vector
${\bf c}=\{c_{\sst{1}},\dots,c_{N}\}$. A \emph{lattice system} over this set of variables is defined by assigning a probability
$\pi_{\bf c}$ to each configuration. The probability distribution $\pi$ satisfies the usual closure condition:
\be\label{PiNormDef1}
\sum_{\bf c} \pi_{\bf c} = 1
\ee
The correlation functions of the system are defined as the average values, built according to $\pi$, of the products of the lattice
variables. It is useful to introduce the $p$-point function $g_{p}$, related to the expectation value of the $c_{i}$'s as follows:
\be\label{Pfunctiondef1}
g_{p}(i_{\sst{1}},\dots ,i_{p}) = \frac{\bra c_{i_{\sst{1}}}\cdots c_{i_{p}} \ket}
{\rho_{i_{\sst{1}}} \cdots \rho_{i_{p}}} \,\, , \quad {\rm where} \quad
\rho_{i} = \bra c_{i} \ket
\ee
where the label $p$ indicates the order of the correlator and ranges from 2 to $N$, while $\bra \phantom{c} \ket$ represents the
statistical average performed with the $\pi$ distribution. The values $\rho_{i}$ of the one-point function represent the (local)
density of the system. In general we will assume that $\rho_{i}$ could depend on the lattice site since we are not restricting to
homogenous systems.

We introduce the notions of entropy, relative entropy and excess entropy that will be widely used in the following. The entropy is
defined, as usual, as the average value of the information associated to the $\pi$ distribution \cite{Shannon1948}, that is:
\be\label{Entropydef1}
S = -\sum_{\bf c} \pi_{\bf c} \ln \pi_{\bf c}
\ee
To define the relative entropy we need to consider a second lattice system over the same set of $c_{i}$ variables. So we introduce
a new distribution, denoted as $\bpi$, that assigns the probability to each configuration. In what follows we will use the symbols
$\pi$ and $\bpi$ to label the two lattice systems. The correlation functions of $\bpi$ are built as those ones of $\pi$ and will
be denoted as $\br_{i}$, $\bg_{\sst{2}}(i,j)$, $\bg_{\sst{3}}(i,j,k),\dots$. The relative entropy between $\pi$ and $\bpi$ is defined
by \cite{Kullback51klDivergence}:
\be\label{RelativeEntropyDef1}
D\bigl(\pi|\bpi\bigr) = \sum_{{\bf c}} \pi_{\bf c}
\ln \frac{\pi_{\bf c}}{\bpi_{\bf c}}
\ee
Equation \eqref{RelativeEntropyDef1} provides a positive number, that vanishes only for identical distributions, which expresses a
quantitative assessment of the statistical distance between the two systems.

Lastly, we introduce the notion of excess entropy for $\pi$ system. This quantity, also called multi-information in the information-theory
oriented literature, expresses the discrepancy between the entropy of $\pi$ and the one of a reference with the same value of the
local density, but uncorrelated otherwise. The excess entropy represents the contribution to the total entropy due to the correlations
among the variables and provides a general measure of the non-independence between the elements of the system
\cite{McGill54,PhysRevLett.91.238701}. Formally, it is defined as:
\be\label{ExcessEntropyExpDef1}
S^{\rm{(ex)}} = - \sum_{\bf c} \pi_{\bf c}^{\sst{(0)}}
\ln \pi_{\bf c}^{\sst{(0)}} +  \sum_{\bf c} \pi_{\bf c} \ln \pi_{\bf c}
\ee
where the reference distribution $\pi_{\bf c}^{\sst{(0)}}$ is the product of the one-point marginals $\pi_{c_{i}}^{\sst{(i)}}$ of
$\pi_{\bf c}$. The factorization properties of $\pi_{\bf c}^{\sst{(0)}}$ allow us to rewrite equation \eqref{ExcessEntropyExpDef1} as:
\be\label{ExcessEntropyExpDef2}
S^{\rm{(ex)}} = \sum_{{\bf c}} \pi_{\bf c}
\ln \frac{\pi_{\bf c}}{\pi_{c_1}^{\sst{(1)}} \cdots \pi_{c_{N}}^{\sst{(N)}}} =
D\bigl(\pi|\pi^{\sst{(0)}} \bigr)
\ee
that is, the relative entropy between the true interacting distribution $\pi$ and the independent model one built by
taking the product of its one-point marginals.

\subsection{Cluster expansion of the excess entropy in a lattice system}
\label{CExcessEntropy}

We collect and discuss some of the results obtained in \cite{springerlink:10.1023/A:1004520432275} for the multiparticle correlation expansion
of the excess entropy on the lattice. For simplicity we restrict our attention to homogenous systems, in which the expectation values of the
one-point function are independent from the lattice site, that is $\bra c_{i} \ket = \rho$.

The excess entropy admits an expansion in terms of correlation functions which is organized as a series in the lattice density
parameter $\rho$:
\be\label{ExcessEntropyExpDef3}
S^{\rm{(ex)}} = \rho^{2} S_{\sst{2}} + \rho^{3} S_{\sst{3}} + \rho^{4} S_{\sst{4}} + \cdots
\ee
An analysis of the results of \cite{springerlink:10.1023/A:1004520432275} evidences that, the generic coefficient $S_{p}$ can be
written as the sum of two different types of contributions, which we refer to as ``regular'' and ``singular'' terms. The former
is expressed as a sum over the $p$-body subspaces, and represents the lattice analogue of the entropy coefficients in the continuum.
The latter are written as sums over subspaces of lower dimensionality with respect to the regular one. These terms have no direct
counterpart in the continuum entropy formula and their presence is due to the singular structure of the lattice. To deal with these
different types of terms we introduce the following notation:
\bea\label{ExcessEntropyExpDef4}
&&S_{\sst{2}} = S_{\sst{2}}^{\sst{(2)}} \nonumber \\
&&S_{\sst{3}} = S_{\sst{3}}^{\sst{(3)}} + S_{\sst{3}}^{\sst{(2)}} \nonumber\\
&&S_{\sst{4}} = S_{\sst{4}}^{\sst{(4)}} + S_{\sst{4}}^{\sst{(3)}} +
S_{\sst{4}}^{\sst{(2)}}  \\
&& \dots \nonumber \\
&&S_{p} = S_{p}^{\sst{(p)}} + S_{p}^{\sst{(p-1)}} + \cdots + S_{p}^{\sst{(2)}} \nonumber
\eea
where the generic coefficient $S_{p}^{(m)}$ indicates the contribution of $m$-dimensional clusters to the $p$-th order in the $\rho$
expansion. So, for instance, the cubic coefficient $S_{\sst{3}}$ is given by the sum of $S_{\sst{3}}^{\sst{(3)}}$ (regular cubic
contribution of three-body clusters) and $S_{\sst{3}}^{\sst{(2)}}$ (singular cubic contribution of the two-body ones).

We report the explicit expressions of the $S_{p}^{(m)}$ coefficients, as they are given in \cite{springerlink:10.1023/A:1004520432275},
up to the third order in $\rho$ and we also present results concerning the fourth order contribution which can be obtained after a long
but straightforward calculation starting from the equation (3.31) of \cite{springerlink:10.1023/A:1004520432275}. Since our definition
of excess entropy \eqref{ExcessEntropyExpDef1} differs from the one  of \cite{springerlink:10.1023/A:1004520432275} for an overall minus
sign, we flip the sign in the formulae of \cite{springerlink:10.1023/A:1004520432275} to provide consistent results.

Firstly, we present the regular contributions that are the lattice analogous of two, three and four body configurational entropy for a
continuous system in the grand canonical ensemble. The second order term reads:
\be\label{S2Entropyclusterexpd1}
S_{\sst{2}}^{\sst{(2)}} = \frac{1}{2} \sum_{i_{1} \neq i_{2}} \biggl(
g_{\sst{2}}(i_{\sst{1}},i_{\sst{2}}) \ln g_{\sst{2}}(i_{\sst{1}},i_{\sst{2}}) - g_{\sst{2}}(i_{\sst{1}},i_{\sst{2}}) + 1 \biggr)
\ee
The third order term is:
\bea\label{S3regEntropyclusterexpd1}
S_{\sst{3}}^{({\sst 3})} &=& \frac{1}{3!} \sum_{i_{1} \neq i_{2} \neq i_{3}} \biggl(
g_{\sst{3}}(i_{\sst{1}},i_{\sst{2}},i_{\sst{3}}) \ln \frac{g_{\sst{3}}(i_{\sst{1}},i_{\sst{2}},i_{\sst{3}})}
{g_{\sst{2}}(i_{\sst{1}},i_{\sst{2}})g_{\sst{2}}(i_{\sst{1}},i_{\sst{3}})g_{\sst{2}}(i_{\sst{2}},i_{\sst{3}})}
+ \biggr. \nonumber \\ &-& \biggl.
g_{\sst{3}}(i_{\sst{1}},i_{\sst{2}},i_{\sst{3}}) +
3g_{\sst{2}}(i_{\sst{1}},i_{\sst{2}})g_{\sst{2}}(i_{\sst{1}},i_{\sst{3}})
- 3g_{\sst{2}}(i_{\sst{1}},i_{\sst{2}}) + 1
\biggr)
\eea
and the four-body regular one reads:
\bea\label{S4regEntropyclusterexpd1}
&& S_{\sst{4}}^{({\sst 4})} = \frac{1}{4!} \sum_{i_{1} \neq i_{2} \neq i_{3} \neq i_{4}}
\Big[  \nonumber \\ &&
g_{\sst{4}}(i_{\sst{1}},i_{\sst{2}},i_{\sst{3}},i_{\sst{4}}) \ln
\frac{g_{\sst{4}}(i_{\sst{1}},i_{\sst{2}},i_{\sst{3}},i_{\sst{4}})
g_{\sst{2}}(i_{\sst{1}},i_{\sst{2}})g_{\sst{2}}(i_{\sst{1}},i_{\sst{3}})
g_{\sst{2}}(i_{\sst{1}},i_{\sst{4}})g_{\sst{2}}(i_{\sst{2}},i_{\sst{3}})
g_{\sst{2}}(i_{\sst{2}},i_{\sst{4}})g_{\sst{2}}(i_{\sst{3}},i_{\sst{4}})}
{g_{\sst{3}}(i_{\sst{1}},i_{\sst{2}},i_{\sst{3}})g_{\sst{3}}(i_{\sst{1}},i_{\sst{2}},i_{\sst{4}})
g_{\sst{3}}(i_{\sst{1}},i_{\sst{3}},i_{\sst{4}})g_{\sst{3}}(i_{\sst{2}},i_{\sst{3}},i_{\sst{4}})}
+ \nonumber \\ &-&
g_{\sst{4}}(i_{\sst{1}},i_{\sst{2}},i_{\sst{3}},i_{\sst{4}}) + 4g_{\sst{3}}(i_{\sst{1}},i_{\sst{2}},i_{\sst{3}})
+4g_{\sst{2}}(i_{\sst{1}},i_{\sst{2}})g_{\sst{2}}(i_{\sst{1}},i_{\sst{3}})g_{\sst{2}}(i_{\sst{1}},i_{\sst{4}})
+ 3g_{\sst{2}}(i_{\sst{1}},i_{\sst{2}})g_{\sst{2}}(i_{\sst{3}},i_{\sst{4}})  + \nonumber \\ &-&
6g_{\sst{2}}(i_{\sst{1}},i_{\sst{2}}) + 6\frac{g_{\sst{3}}(i_{\sst{1}},i_{\sst{2}},i_{\sst{3}})
g_{\sst{3}}(i_{\sst{1}},i_{\sst{2}},i_{\sst{4}})}{g_{\sst{2}}(i_{\sst{1}},i_{\sst{2}})}
- 12g_{\sst{2}}(i_{\sst{1}},i_{\sst{2}})g_{\sst{3}}(i_{\sst{1}},i_{\sst{3}},i_{\sst{4}}) + 2 \Big]
\eea
Then we collect the expression of the singular terms up to the fourth order. Two-body clusters contribute at both the third and
fourth order in $\rho$. The cubic one is:
\be\label{S3singEntropyclusterexpd1}
S_{\sst{3}}^{({\sst 2})} = \frac{1}{2} \sum_{i_{1} \neq i_{2}} \biggl(
g_{\sst{2}}(i_{\sst{1}},i_{\sst{2}}) - 1
\biggr)^{2}
\ee
and the quartic one reads:
\be\label{S4sing2Entropyclusterexpd1}
S_{\sst{4}}^{({\sst 2})} = \frac{1}{2} \sum_{i_{1} \neq i_{2}} \biggl(
g_{\sst{2}}(i_{\sst{1}},i_{\sst{2}}) - 1 \biggr)^{2}
\left(\frac{1}{3}g_{\sst{2}}(i_{\sst{1}},i_{\sst{2}}) + \frac{7}{6}\right)
\ee
Lastly, we write the three-body singular term that appears at the fourth order in $\rho$:
\bea\label{S4sing3Entropyclusterexpd1}
S_{\sst{4}}^{({\sst 3})} &=& \frac{1}{3!} \sum_{i_{1} \neq i_{2} \neq i_{3}} \left[
\frac{3}{2}\frac{g_{\sst{3}}^{2}(i_{\sst{1}},i_{\sst{2}},i_{\sst{3}})}{g_{\sst{2}}(i_{\sst{1}},i_{\sst{2}})}
+ 3g_{\sst{3}}(i_{\sst{1}},i_{\sst{2}},i_{\sst{3}})
- 6g_{\sst{2}}(i_{\sst{1}},i_{\sst{2}})g_{\sst{3}}(i_{\sst{1}},i_{\sst{2}},i_{\sst{3}}) +
\right. \nonumber \\  &+& \left.
3g_{\sst{2}}^{2}(i_{\sst{1}},i_{\sst{2}})g_{\sst{2}}(i_{\sst{1}},i_{\sst{3}}) +
3g_{\sst{2}}(i_{\sst{1}},i_{\sst{2}})g_{\sst{2}}(i_{\sst{1}},i_{\sst{3}})
- \frac{15}{2}g_{\sst{2}}(i_{\sst{1}},i_{\sst{2}}) + 3 \right]
\eea

\section{Multiparticle correlation expansion of the relative entropy}
\label{CERelativeEntropy}

As stated in section \ref{PreliminaryResults} we consider two lattice systems, denoted as $\pi$ and $\bpi$, both defined over the same
set of $N$ discrete variables. According to its standard definition, the relative entropy $D(\pi|\bpi)$ between $\pi$ and $\bpi$ is given
by equation \eqref{RelativeEntropyDef1}, and its assessment involves a sum over the whole $2^{N}$-dimensional configuration space. The
aim of our analysis is to recast this sum into a different form, in which each addend can be ascribed to a specific cluster of lattice sites.
This procedure provides an equivalent reformulation of the RE when all possible clusters are included. Nonetheless, pursuing this approach
allows us to define a family of functions, denoted as ${\cal D}_{p}$, which constitutes an approximation of the complete RE, limited to the
contribution of clusters of maximum dimension equal to $p<N$.

The explicit derivation of ${\cal D}_{p}$ constitutes the main goal of our analysis. To achieve this result we start by introducing the main
building block of this construction, namely the cluster relative entropy ${\cal C}_{k}$. We define a generic $k$-dimensional cluster as a
subset of the configuration space made of the variables $c_{i_{1}},\dots,c_{i_{k}}$ and introduce the notion of cluster relative entropy
as follows:
\be\label{CREDef1}
{\cal C}_{k}(i_{\sst{1}},\dots,i_{k}) = \sum_{c_{i_{1}},\dots,c_{i_{k}}} \pi_{c_{i_{1}}\cdots c_{i_{k}}}
\ln \frac{\pi_{c_{i_{1}}\cdots c_{i_{k}}}}{\bpi_{c_{i_{1}}\cdots c_{i_{k}}}}
\ee
where the cluster probabilities $\pi_{c_{i_{1}}\cdots c_{i_{k}}}$, $\bpi_{c_{i_{1}}\cdots c_{i_{k}}}$ are obtained from those one of
$\pi$ and $\bpi$ through a marginalization of the variables that do not belong to the cluster.

Then, in order to express the ${\cal D}_{p}$'s in term of the cluster relative entropies, we perform a factorization of the ratio $\pi_{\bf c}
/\bpi_{\bf c}$ analogous to the one already introduced in the literature connected to the cluster expansion of the configurational entropy.
The details of this procedure are described in appendix \ref{appendixA}. Pursuing this approach provides:
\be\label{DpDef1}
{\cal D}_{p} = \sum_{k=1}^{p} \a_{k}\sum_{i_{\sst{1}}<\cdots<i_{k}} {\cal C}_{k}(i_{\sst{1}},\dots,i_{k})
\ee
Equation \eqref{DpDef1} states that ${\cal D}_{p}$ is expressed as the sum of the relative entropies associated to all the distinct clusters
with dimension ranging from 1 to $p$. The $\a_{k}$ coefficients represent the multiplicity factors that weight the contribution of the different
orders. Since we are interested in writing a multiparticle correlator expansion of the RE we need to express  ${\cal C}_{k}(i_{\sst{1}},
\dots,i_{k})$ in terms of the correlation functions of $\pi$ and $\bpi$. So, to complete the achievement of our main result, we have to work out
the general expression of the cluster relative entropies and to compute the $\a_{k}$ coefficients. Both of these topics will be analyzed in the
next two subsections.

\subsection{Expression of the cluster relative entropy}
\label{ClusterRelEntropy}

The aim of this subsection is to express the clusters probabilities $\pi_{c_{i_{1}}\cdots c_{i_{k}}}$ and $\bpi_{c_{i_{1}}\cdots c_{i_{k}}}$
in terms of the correlation functions of $\pi$ and $\bpi$, respectively. Once that this is done, the multiparticle correlator formulation
of the cluster relative entropy can be readily obtained from equation \eqref{CREDef1}. We start by building an explicit solution for low
dimensional clusters and then we write the generic equations valid for the arbitrary $k$-dimensional case.

\subsubsection*{One-dimensional cluster}

The cluster has 2 states, which correspond to the values of the (only) lattice variable $c_{i}$. The distribution function of the system
is described in terms of the parameter $\pi_{\sst{1}}$, which represents the probability associated of the configuration $c_{i} = 1$.
The second parameter $\pi_{\sst{0}}$, that expresses the probability of the configurations $c_{i} = 0$, is determined by the closure
condition, so that $\pi_{\sst{0}} = 1 - \pi_{\sst{1}}$.
We impose that the one-point function $\bra c_{i} \ket$ is equal to $\rho_{i}$. This implies:
\be\label{OnePointPiDef1}
\bra c_{i} \ket = \pi_{\sst{1}} = \rho_{i}
\ee
These conditions allow us to univocally parametrize $\pi_{\sst{1}}$ and $\pi_{\sst{0}}$ in terms of the one-point function. Indeed:
\be\label{OnePointPiDef2}
\pi_{\sst{1}} = \rho_{i} \,\, , \qquad
\pi_{\sst{0}} = 1 - \rho_{i}
\ee

\subsubsection*{Two-dimensional cluster}

The cluster $\{c_{i_{1}},c_{i_{2}}\}$ possesses 4 states represented by the possible values of its couple of lattice variables. The elements
of the distribution function $\pi_{c_{i_{1}}c_{i_{2}}}$ read $\pi_{\sst{11}}$, $\pi_{\sst{10}}$, $\pi_{\sst{01}}$, $\pi_{\sst{00}}$ and the closure
condition holds, so that $\pi_{\sst{00}} = 1 - \pi_{\sst{11}} - \pi_{\sst{10}} - \pi_{\sst{01}}$.
In this case we impose the values of the one and two-point correlation functions, namely:
\bea\label{TwoPointPiDef1}
&&\bra c_{i_{1}} \ket = \pi_{\sst{11}} + \pi_{\sst{10}} = \rho_{i_{1}} \qquad
\bra c_{i_{2}} \ket = \pi_{\sst{11}} + \pi_{\sst{01}} = \rho_{i_{2}} \nonumber \\
&& \bra c_{i_{1}} c_{i_{2}} \ket = \pi_{\sst{11}} = \rho_{i_{1}} \rho_{i_{2}} g_{\sst{2}}(i_{\sst{1}},i_{\sst{2}})
\eea
These equations can be solved and, together with the closure condition, provide the expression of the $\pi_{c_{i_{1}}c_{i_{2}}}$ parameters in terms
of the cluster correlation functions:
\bea\label{TwoPointPiDef2}
&&\pi_{\sst{11}} = \rho_{i_{\sst{1}}} \rho_{i_{\sst{2}}} g_{\sst{2}}(i_{\sst{1}},i_{\sst{2}}) \nonumber \\
&&\pi_{\sst{10}} = \rho_{i_{\sst{1}}} - \rho_{i_{\sst{1}}} \rho_{i_{\sst{2}}} g_{\sst{2}}(i_{\sst{1}},i_{\sst{2}}) \nonumber \\
&&\pi_{\sst{01}} = \rho_{i_{\sst{2}}} - \rho_{i_{\sst{1}}} \rho_{i_{\sst{2}}} g_{\sst{2}}(i_{\sst{1}},i_{\sst{2}}) \\
&&\pi_{\sst{00}} = 1 - \rho_{i_{\sst{1}}} - \rho_{i_{\sst{2}}} + \rho_{i_{\sst{1}}} \rho_{i_{\sst{2}}} g_{\sst{2}}(i_{\sst{1}},i_{\sst{2}})
\nonumber
\eea

\subsubsection*{$k$-dimensional cluster}

An analysis of equations \eqref{TwoPointPiDef1} evidences that $\pi_{\sst{11}}$, associated to the configuration in which both the lattice
variables are equal to 1, is readily given in term of the two-point function. Instead, $\pi_{\sst{10}}$ and $\pi_{\sst{01}}$ are written by
substituting the expression of $\pi_{\sst{11}}$ in the equations for the one-point function. Finally, $\pi_{\sst{00}}$ is obtained from the
closure condition.

The same type of hierarchical structure holds also in the generic case and it is possible to write the solution for the $\pi$ parameters of
a $k$-dimensional cluster by generalizing the relations \eqref{TwoPointPiDef2}. To this scope, we classify the configurations of the cluster
according to the number $q$ of lattice variables equal to 1 inside the given configuration ($q$ ranges from 0 to $k$). In order to present the
general solution it is useful to introduce three different types of multi-index labels. The first one, $i^{(k)}=i_{\sst{1}},\ldots,i_{k}$ identifies
the $k$ sites of the cluster. The second one, $j^{(q)}=j_{\sst{1}},\ldots,j_{q}$ labels the $q$ sites of the cluster for which $c=1$ in the given
configuration. The third one, $l^{(r)}=l_{\sst{1}},\ldots,l_{r}$ describes a generic subset of the cluster made of $r$ lattice sites. For each
configuration $\{c_{i_{1}},\dots,c_{i_{k}}\}$ we affirm that $\pi_{c_{i_{1}}\cdots c_{i_{k}}}$ can be written in terms of the cluster correlation
functions according to the following relations:
\bea\label{kPointPiDef1}
&&\pi_{c_{i_{1}} \cdots c_{i_{k}}} = \sum_{r=0}^{k-q}\frac{(-)^{r}}{r!}
\sum_{l^{(r)} \neq j^{(q)}} \r_{j_{1}}\dots\r_{j_{q}}\r_{l_{1}}\dots\r_{l_{r}}
g_{q+r}\left(j^{(q)},l^{(r)}\right) \,\,\, , \,\,\, {\rm if} \,\,\, q>0
\nonumber \\
&&\pi_{\sst{0}\dots\sst{0}} = 1 + \sum_{r=1}^{k}(-)^{r}\sum_{l_{\sst{1}} < \cdots < l_{r}}
\r_{l_{1}}\dots\r_{l_{r}}g_{r}\left(l^{(r)}\right) \,\,\, , \,\,\, {\rm if} \,\,\, q=0
\eea
where we have implicitly defined $g_{\sst{1}}(i) = 1$ and we remember that $g_{p}$ is a symmetric functions of all its arguments.
Formulae \eqref{kPointPiDef1} hold for cluster of arbitrary dimension and we can verify their correctness for low values of $k$. First of all,
it is immediate to check that relations \eqref{OnePointPiDef2} and \eqref{TwoPointPiDef2} are reproduced for $k=1$ and 2, respectively. As a
further example, just to go beyond the two-dimensional case, we apply \eqref{kPointPiDef1} to the cluster $\{c_{i_{1}},c_{i_{2}},c_{i_{3}}\}$
with $k=3$:
\bea\label{ThreePointPiDef1}
&&\pi_{\sst{111}} = \rho_{i_{1}} \rho_{i_{2}} \rho_{i_{3}} g_{\sst{3}}(i_{\sst{1}},i_{\sst{2}},i_{\sst{3}}) \nonumber \\
&&\pi_{\sst{110}} = \rho_{i_{1}} \rho_{i_{2}} g_{\sst{2}}(i_{\sst{1}},i_{\sst{2}})  -
\rho_{i_{1}} \rho_{i_{2}} \rho_{i_{3}} g_{\sst{3}}(i_{\sst{1}},i_{\sst{2}},i_{\sst{3}}) \nonumber \\
&&\pi_{\sst{101}} = \rho_{i_{1}} \rho_{i_{3}} g_{\sst{2}}(i_{\sst{1}},i_{\sst{3}})  -
\rho_{i_{1}} \rho_{i_{2}} \rho_{i_{3}} g_{\sst{3}}(i_{\sst{1}},i_{\sst{2}},i_{\sst{3}}) \nonumber \\
&&\pi_{\sst{011}} = \rho_{i_{2}} \rho_{i_{3}} g_{\sst{2}}(i_{\sst{2}},i_{\sst{3}})  -
\rho_{\sst{1}} \rho_{i_{2}} \rho_{i_{3}} g_{\sst{3}}(i_{\sst{1}},i_{\sst{2}},i_{\sst{3}}) \nonumber \\
&&\pi_{\sst{100}} = \rho_{i_{1}} - \rho_{i_{1}} \rho_{i_{2}} g_{\sst{2}}(i_{\sst{1}},i_{\sst{2}})
- \rho_{i_{1}} \rho_{i_{3}} g_{\sst{2}}(i_{\sst{1}},i_{\sst{3}}) +
\rho_{i_{1}} \rho_{i_{2}} \rho_{i_{3}} g_{\sst{3}}(i_{\sst{1}},i_{\sst{2}},i_{\sst{3}}) \\
&&\pi_{\sst{010}} = \rho_{i_{2}} - \rho_{i_{1}} \rho_{i_{2}} g_{\sst{2}}(i_{\sst{1}},i_{\sst{2}})
- \rho_{i_{2}} \rho_{i_{3}} g_{\sst{2}}(i_{\sst{2}},i_{\sst{3}}) +
\rho_{i_{1}} \rho_{i_{2}} \rho_{i_{3}} g_{\sst{3}}(i_{\sst{1}},i_{\sst{2}},i_{\sst{3}}) \nonumber \\
&&\pi_{\sst{001}} = \rho_{i_{3}} - \rho_{i_{1}} \rho_{i_{3}} g_{\sst{2}}(i_{\sst{1}},i_{\sst{3}})
- \rho_{i_{2}} \rho_{i_{3}} g_{\sst{2}}(i_{\sst{2}},i_{\sst{3}}) +
\rho_{i_{1}} \rho_{i_{2}} \rho_{i_{3}} g_{\sst{3}}(i_{\sst{1}},i_{\sst{2}},i_{\sst{3}}) \nonumber \\
&&\pi_{\sst{000}} = 1 - \rho_{i_{1}} - \rho_{i_{2}} - \rho_{i_{3}} +
\rho_{i_{1}} \rho_{i_{2}} g_{\sst{2}}(i_{\sst{1}},i_{\sst{2}}) +
\rho_{i_{1}} \rho_{i_{3}} g_{\sst{2}}(i_{\sst{1}},i_{\sst{3}}) +
\rho_{i_{2}} \rho_{i_{3}} g_{\sst{2}}(i_{\sst{2}},i_{\sst{3}}) + \nonumber \\ &&\hspace{1cm}
-\rho_{i_{1}} \rho_{i_{2}} \rho_{i_{3}} g_{\sst{3}}(i_{\sst{1}},i_{\sst{2}},i_{\sst{3}}) \nonumber
\eea
and it is easy to check that the $\pi$ written in this way represent a normalized probability distribution with the expected values
of the one, two and three points correlation functions.

\subsubsection*{Cluster relative entropy}

In the previous subsection we have expressed the cluster probabilities $\pi_{c_{i_{1}}\cdots c_{i_{k}}}$ in terms of the correlation
functions of $\pi$. Obviously, the same procedure can be developed also for the $\bpi$ system and gives a solutions, formally
identical to \eqref{kPointPiDef1}, in which the correlator of $\pi$ are substituted by the ones of $\bpi$. Using these quantities as
input, we can compute the cluster relative entropy directly from its definition \eqref{CREDef1}. For a sake of concreteness we present
some low $k$ examples.
For a one-dimensional cluster, formulae \eqref{OnePointPiDef2} give:
\be\label{C1def1}
{\cal C}_{\sst{1}}(i) = \rho_{i} \ln \frac{\rho_{i}}{\br_{i}} + \left(1-\rho_{i}\right)
\ln \frac{1-\rho_{i}}{1-\br_{i}}
\ee
While, for $k=2$, the expressions \eqref{TwoPointPiDef2} provide:
\bea\label{C2def1}
{\cal C}_{\sst{2}}(i_{\sst{1}},i_{\sst{2}}) &=&  \rho_{i_{1}}\rho_{i_{2}}g_{\sst{2}}(i_{\sst{1}},i_{\sst{2}})
\ln\frac{\rho_{i_{1}}\rho_{i_{2}}g_{\sst{2}}(i_{\sst{1}},i_{\sst{2}})}{\br_{i_{1}}\br_{i_{2}}\bg_{\sst{2}}(i_{\sst{1}},i_{\sst{2}})} +
\left(\rho_{i_{1}}-\rho_{i_{1}}\rho_{i_{2}}g_{\sst{2}}(i_{\sst{1}},i_{\sst{2}})\right)\ln
\frac{\rho_{i_{1}}-\rho_{i_{1}}\rho_{i_{2}}g_{\sst{2}}(i_{\sst{1}},i_{\sst{2}})}
{\br_{i_{1}}-\br_{i_{1}}\br_{i_{2}}\bg_{\sst{2}}(i_{\sst{1}},i_{\sst{2}})}+
\nonumber \\ &+&
\left(\rho_{i_{2}}-\rho_{i_{1}}\rho_{i_{2}}g_{\sst{2}}(i_{\sst{1}},i_{\sst{2}})\right)\ln
\frac{\rho_{i_{2}}-\rho_{i_{1}}\rho_{i_{2}}g_{\sst{2}}(i_{\sst{1}},i_{\sst{2}})}
{\br_{i_{2}}-\br_{i_{1}}\br_{i_{2}}\bg_{\sst{2}}(i_{\sst{1}},i_{\sst{2}})}+
\nonumber \\ &+&
\left(1-\rho_{i_{1}}-\rho_{i_{2}} +\rho_{i_{1}}\rho_{i_{2}}g_{\sst{2}}(i_{\sst{1}},i_{\sst{2}})\right)\ln
\frac{1-\rho_{i_{1}}-\rho_{i_{2}} +\rho_{i_{1}}\rho_{i_{2}}g_{\sst{2}}(i_{\sst{1}},i_{\sst{2}})}
{1-\br_{i_{1}}-\br_{i_{2}} +\br_{i_{1}}\br_{i_{2}}\bg_{\sst{2}}(i_{\sst{1}},i_{\sst{2}})}
\eea
In the same way we can use relations \eqref{ThreePointPiDef1}, and the analogous ones for the $\bpi$ system, and write the
cluster relative entropy at the level $k=3$ by summing over the eight states of the three dimensional clusters. That is:
\be\label{C3def1}
{\cal C}_{\sst{3}}(i_{\sst{1}},i_{\sst{2}},i_{\sst{3}}) =
\sum_{c_{i_{1}},c_{i_{2}},c_{i_{3}}} \pi_{c_{i_{1}}c_{i_{2}}c_{i_{3}}}
\ln \frac{\pi_{c_{i_{1}}c_{i_{2}}c_{i_{3}}}}{\bpi_{c_{i_{1}}c_{i_{2}}c_{i_{3}}}}
\ee

\subsection{Computation of the $\a_{k}$ coefficients}
\label{alphaCoefficients}

As explained in the appendix \ref{appendixA}, the $\a_{k}$ coefficients represent the multiplicity factors associated to the $k$-dimensional
clusters when the family of ${\cal D}_{p}$ functions, formally defined by equation \eqref{FactorizationDef3}, are expressed in terms of the
cluster relative entropies, as stated in \eqref{DpDef1}.

In order to evaluate these coefficients, we count the multiplicity of each cluster, as a function of its dimensionality, starting from the highest
value, that is $k=p$. In this case, since \eqref{FactorizationDef3} contains an explicit sum over $i_{\sst{1}}<\cdots<i_{p}$, all the distinct
$p$-dimensional clusters that can be built starting from the $N$ lattice variables are counted one time, so we immediately conclude that:
\be
\a_{p}=1 \nonumber
\ee
The computation becomes more involved at the level $k=p-1$, where we have to count the number of occurrences of clusters of the type
$\{c_{i_{1}},c_{i_{2}},\dots,c_{i_{p-1}}\}$. This object contributes in two distinct points when the relation \eqref{FactorizationDef4} is plugged
into \eqref{FactorizationDef3}. Firstly, it appears, with a multiplicity equal to one, as the leading order contribution at the $(p-1)$ level.
Secondly, it also emerges as a subset of all the $p$-dimensional clusters of the type $\{c_{i_{1}},c_{i_{2}},\dots,c_{i_{p-1}},c_{l}\}$.
Here $l$ represents one of the $N-(p-1)$ possible choices of the lattice site index that are not included in the $(p-1)$-dimensional cluster.
So, the multiplicity factor at the level $(p-1)$ is given by:
\be
\a_{p-1} = 1 - (N-(p-1)) = -(N-p)
\nonumber
\ee
Going to the next level, that is $k=p-2$, presents an analogous situation but in this case there are contributions coming from higher dimensional
clusters of both orders $p-1$ and $p$. A counting of the multiplicity analogous to one previously described allows us to write the following
equation for $\a_{p-2}$:
\be
\a_{p-2} = 1 -(N-(p-2))+\frac{1}{2}(N-(p-2))(N-(p-1)) \nonumber
\ee
which can be written as:
\be
\a_{p-2} = \frac{1}{2}(N-p)(N-p+1) \nonumber
\ee
The computation of the coefficients associated to clusters of lower dimensionality proceeds by iterating this type of combinatorial procedure.
At the generic $k$-level one has:
\be
\alpha_{k} = \sum_{i=0}^{p-k} (-)^{i} \binom{N-k}{i} \nonumber
\ee
where the binomial coefficient expresses the number of $(k+i)$-dimensional clusters that contain a specific $k$-dimensional one as a subset. The
sum, taken with alternate sign according to \eqref{FactorizationDef4}, is extended up to $i=p-k$ in order to include clusters of maximum dimension
equal to $p$. Then, by making usage of the standard algebraic properties of binomial coefficients \cite{Abramowitz:1964:HMF}:
\be
\binom{N}{k} = \binom{N}{N-k} \,\,\, , \quad
\sum_{i=0}^{k} (-)^{i} \binom{N}{i} = (-)^{k}\binom{N-1}{k} \nonumber
\ee
we can write $\alpha_{k}$ as:
\be\label{alphakDef1}
\alpha_{k} = (-)^{p-k} \binom{N-k-1}{N-p-1} = \frac{(-)^{p-k}}{(p-k)!} \prod_{i=k+1}^{p}(N-i)
%\,\,\, , \,\,\, \textrm{for} \,\,\,k=1,\dots,p-1
\ee
where the last expression is valid for values of $k$ from 1 up to $p-1$, while $\a_{p}=1$ for the highest dimensional clusters.

\subsection{General properties of ${\cal D}_{p}$ and some explicit results}
\label{GeneralProperties}

Thanks to the results previously achieved we are now ready to use equation \eqref{DpDef1} to write the explicit expression of the multiparticle
correlation expansion of the RE limited to clusters of maximum dimension equal to $p$. Before presenting some explicit results for low values of $k$
we analyze the extreme case, that is $p=N$. In this case, the sum over clusters of \eqref{DpDef1} is extended up to the $N$-dimensional element
that coincides with the complete lattice system. Consequently, we expect that the corresponding ${\cal D}_{N}$ function should reproduce the complete
relative entropy \eqref{RelativeEntropyDef1}.
The analysis of equation \eqref{DpDef1} for $p=N$ confirms that this is indeed the case. In effect, an inspection of \eqref{alphakDef1} evidences that,
apart from $\a_{p}$ that is always equal to one, all the $\a_{k}$ coefficients contain a $(N-p)$ factor. So, when $p=N$ only $\a_{N}$ is non vanishing
and consequently  ${\cal D}_{N}$ is equal to ${\cal C}_{N}$ that, according to its definition \eqref{CREDef1}, is the complete relative entropy between
$\pi$ and $\bpi$.

We now discuss the application of equation \eqref{DpDef1} to derive some explicit results for low values of $p$. The expressions of the cluster
relative entropies up to $p=3$ are given in the end of the section \ref{ClusterRelEntropy}. So, using \eqref{alphakDef1} to compute the $\a_{k}$
coefficients for each value of $p=1,2,3$ we have:
\bea\label{D1-3def1}
&& {\cal D}_{\sst{1}} = \sum_{i} {\cal C}_{\sst{1}}(i) \nonumber \\
&& {\cal D}_{\sst{2}} = \sum_{i_{1}<i_{2}} {\cal C}_{\sst{2}}(i_{\sst{1}},i_{\sst{2}})
- (N-2) \sum_{i} {\cal C}_{\sst{1}}(i) \\
&& {\cal D}_{\sst{3}} = \sum_{i_{1}<i_{2}<i_{3}} {\cal C}_{\sst{3}}(i_{\sst{1}},i_{\sst{2}},i_{\sst{3}})
-(N-3) \sum_{i_{1}<i_{2}} {\cal C}_{\sst{2}}(i_{\sst{1}},i_{\sst{2}})
+ \frac{1}{2}(N-3)(N-2) \sum_{i} {\cal C}_{\sst{1}}(i) \nonumber
\eea
A first comment that emerges looking at the combinatorial structure of \eqref{D1-3def1} concerns with the sign of the ${\cal D}_{p}$ functions.
Indeed, the cluster relative entropies \eqref{CREDef1} are always non negative but the $\a_{k}$ coefficients contain a $(-)^{p-k}$ factor, so it is
not immediately evident if the linear combinations \eqref{D1-3def1} possess a definite sign. Nonetheless, it is possible to show that each r.h.s of
\eqref{D1-3def1} is always non negative. The proof of this statement is based on the same relation $\ln x \leq x-1$ that is used to prove the non
negativity of the complete relative entropy \cite{Kullback51klDivergence}. In this case it is convenient to resort to the expression of ${\cal D}_{p}$
given by \eqref{FactorizationDef3} and to expand the logarithm of the product of the $R$ ratios in terms of the sum of the logarithms of the single
factors. Then, by applying the afore mentioned inequality, it is easy to verify that each addend of the r.h.s of \eqref{FactorizationDef3} is always
greater than zero (and vanishes only if the cluster probabilities of $\pi$ and $\bpi$ are equal). This argument is completely general and can be
directly extended to all the values of $p$.

A last interesting comment regards the dependence of the present results from the density of the systems. According to \eqref{DpDef1}, the ${\cal D}_{p}$
functions are written as linear combinations of the cluster relative entropies ${\cal C}_{k}$ that, in turn, are computed on the base of equations
\eqref{kPointPiDef1}, which relate cluster probabilities to correlation functions. So, none of the steps needed for the achievement of our procedures requires
an explicit expansion in powers of the density. Consequently the ${\cal D}_{p}$'s are non polynomial in $\rho$ and equations \eqref{D1-3def1} represent an
example of a multiparticle correlation expansion that is not explicitly written as a series in powers of the density. Obviously, this observation is not
in conflict with the general role of the density as a control parameter for the convergence of a cluster expansion and we expect that the reliability
of the assessment of the true relative entropy based on a given ${\cal D}_{p}$ increases at low density.

\section{Novel derivation of the excess entropy cluster expansion}
\label{EntropyCE}

As a first application of the cluster expansion of the relative entropy we propose a novel derivation of the multiparticle correlation expansion
of the excess entropy on the lattice. The connection between excess entropy and relative entropy has been established in section \ref{BasicNotations}.
In particular, equation \eqref{ExcessEntropyExpDef2} states that the excess entropy of the $\pi$ system can be expressed as the RE between its probability
distribution and the independent-model one $\pi^{\sst{(0)}}$, built as the product of the one-point marginals of $\pi$. So, starting from this correspondence,
we can express the multiparticle correlation expansion of the excess entropy by specifying the general formulae for ${\cal D}_{p}$, derived in the previous
section, to this specific choice of the $\bpi$ distribution.

For simplicity we restrict our analysis to homogeneous systems, for which the value of the one-point function is independent from the lattice
site. So, the expression of the cluster probabilities $\pi_{c_{i_{1}}\cdots c_{i_{k}}}$ are obtained by setting $\r_{i}=\r$  in the formulae
\eqref{kPointPiDef1}. Moreover, $\pi^{\sst{(0)}}$ represents a system with the same density of $\pi$ but uncorrelated at the level of higher-point
functions. Consequently, its cluster probabilities $\pi^{\sst{(0)}}_{c_{i_{1}}\cdots c_{i_{k}}}$ are obtained by the ones of $\pi$ by setting $g_{k}=1$.
Their expression can be written in the form:
\bea\label{kPointPiZeroHomDef1}
\pi^{\sst{(0)}}_{c_{i_{1}} \cdots c_{i_{k}}} = \sum_{r=0}^{k-q}(-)^{r}\binom{k-q}{r}
\r^{q+r} \,\,\, , \,\,\, q = 0,\dots,k
\eea
where, as usual, $q$ identifies the number of lattice variables equal to 1 in the configuration $\{c_{i_{1}},\dots,c_{i_{k}}\}$.

We are now ready to compute the ${\cal D}^{\sst{(0)}}_{p}$ functions, obtained by substituting the cluster probabilities defined above in formula
\eqref{DpDef1}. An inspection of equation \eqref{C1def1} evidences that the contribution of the one-dimensional clusters vanishes, since the
value of the one-point function associated to the reference $\pi^{\sst{(0)}}$ distribution coincides with the one of $\pi$. The first non trivial
contribution comes from clusters of dimension two and reads:
\bea\label{D02pointdef1}
{\cal D}^{\sst{(0)}}_{\sst{2}} &=&  \sum_{i_{1} < i_{2}}\left[\rho^{2}g_{\sst{2}}(i_{\sst{1}},i_{\sst{2}})
\ln g_{\sst{2}}(i_{\sst{1}},i_{\sst{2}}) + 2\left(\rho-\rho^{2}g_{\sst{2}}(i_{\sst{1}},i_{\sst{2}})\right)\ln
\frac{\rho-\rho^{2}g_{\sst{2}}(i_{\sst{1}},i_{\sst{2}})}{\rho-\rho^{2}}+
\right. \nonumber \\ &+& \left.
\left(1-2\rho +\rho^{2}g_{\sst{2}}(i_{\sst{1}},i_{\sst{2}})\right)\ln
\frac{1-2\rho +\rho^{2}g_{\sst{2}}(i_{\sst{1}},i_{\sst{2}})}{1-2\rho+\rho^{2}}
\right]
\eea
Equation \eqref{D02pointdef1} represents, by definition, the contribution of two dimensional clusters to the excess entropy of the $\pi$ system.
Looking at this formula we observe that its is inherently non polynomial in the lattice density and, in order to compare this result with the
standard expansion of the excess entropy, presented in \cite{springerlink:10.1023/A:1004520432275}, we perform a Taylor expansion in $\rho$.
Results up to the fourth order read:
\be\label{D02pointdef2}
{\cal D}^{\sst{(0)}}_{\sst{2}}\Big{|}_{\rho} = 0 \,\, , \quad
{\cal D}^{\sst{(0)}}_{\sst{2}}\Big{|}_{\rho^{2}} = S_{\sst{2}} \,\, , \quad
{\cal D}^{\sst{(0)}}_{\sst{2}}\Big{|}_{\rho^{3}} = S_{\sst{3}}^{(\sst{2})} \,\, , \quad
{\cal D}^{\sst{(0)}}_{\sst{2}}\Big{|}_{\rho^{4}} = S_{\sst{4}}^{(\sst{2})}
\ee
where $S_{\sst{2}},S_{\sst{3}}^{({\sst 2})},S_{\sst{4}}^{({\sst 2})}$ have been defined in equation \eqref{ExcessEntropyExpDef4} of section
\ref{CExcessEntropy}. So, the coefficient extracted from the density power expansion of ${\cal D}^{\sst{(0)}}_{\sst{2}}$ at the quadratic order
reproduces the regular two-body term \eqref{S2Entropyclusterexpd1}, the one coming from the cubic order provides the singular two-body cubic term
\eqref{S3singEntropyclusterexpd1}, while the fourth order coefficient is equal to singular two-body quartic term \eqref{S4sing2Entropyclusterexpd1}.
This result constitutes an example of what stated in the last paragraph of the previous section: ${\cal D}^{\sst{(0)}}_{\sst{2}}$ represents
a multiparticle correlation expansion which is not explicitly written as a series in powers of the density. Indeed, results collected in \eqref{D02pointdef2}
confirm that we can interpret it as a (partial) resummation of the series \eqref{ExcessEntropyExpDef3}, limited to the terms associated to
clusters of dimension two.

We now perform the same type of analysis at the level $p=3$,  where all the clusters up to the three dimensional ones are taken into account. We
compute the function ${\cal D}^{\sst{(0)}}_{\sst{3}}$ starting from the third line of \eqref{D1-3def1}. This provides:
\be\label{D03pointdef1}
{\cal D}^{\sst{(0)}}_{\sst{3}} =  \sum_{i_{1} < i_{2} < i_{3}}
{\cal C}^{\sst{(0)}}_{\sst{3}}(i_{\sst{1}},i_{\sst{2}},i_{\sst{3}})
-(N-3) {\cal D}^{\sst{(0)}}_{\sst{2}}
\ee
where ${\cal C}^{\sst{(0)}}_{\sst{3}}(i_{\sst{1}},i_{\sst{2}},i_{\sst{3}})$ is computed starting from equation \eqref{C3def1}. Here the $\pi$'s are
obtained from equations \eqref{ThreePointPiDef1} with $\r_{i}=\r$, and the $\bpi$'s are built according to \eqref{kPointPiZeroHomDef1} with $k=3$.
As in the previous case the functional structure of ${\cal D}^{\sst{(0)}}_{\sst{3}}$ is non polynomial in the density and the same type of considerations
made for $p=2$ hold in this case too. Indeed, performing a Taylor expansion in $\r$ up to the fourth order provides:
\be\label{D03pointdef2}
{\cal D}^{\sst{(0)}}_{\sst{3}}\Big{|}_{\rho} = 0 \,\, , \quad
{\cal D}^{\sst{(0)}}_{\sst{3}}\Big{|}_{\rho^{2}} = S_{\sst{2}} \,\, , \quad
{\cal D}^{\sst{(0)}}_{\sst{3}}\Big{|}_{\rho^{3}} = S_{\sst{3}}^{({\sst 3})} + S_{\sst{3}}^{({\sst 2})} \,\, , \quad
{\cal D}^{\sst{(0)}}_{\sst{3}}\Big{|}_{\rho^{4}} = S_{\sst{4}}^{({\sst 3})} + S_{\sst{4}}^{({\sst 2})}
\ee
Here we recognize that the density expansion of ${\cal D}^{\sst{(0)}}_{\sst{3}}$ reproduce the expected values of the regular and singular terms
of \cite{springerlink:10.1023/A:1004520432275}, due to clusters of dimensions two and three.

As a further check we have built the ${\cal D}_{\sst{4}}^{\sst{(0)}}$ function, starting from the general equation \eqref{DpDef1}, and we have been able
to derive all the coefficients of equation \eqref{ExcessEntropyExpDef4}, up to the terms due to clusters of dimension four. These results provide a further
confirmation, at the level $p=3$ and $p=4$, of the afore mentioned interpretation of ${\cal D}^{\sst{(0)}}_{p}$ as the resummation of the series
\eqref{ExcessEntropyExpDef3} limited to the contribution of clusters of maximum dimension equal to $p$.

\section{Discussion and conclusions}
\label{Conclusions}

In this paper we have discussed the construction of the multiparticle correlation expansion of relative entropy for lattice systems. The
main result of the present analysis is provided by the definition of the ${\cal D}_{p}$ functions, that express the assessment of the relative
entropy based on a multiparticle correlation expansion limited to clusters of maximum dimension equal to $p$. We have presented a systematic procedure
to calculate these functions, for arbitrary values of $p$. Our analysis has employed a standard cluster factorization formula for the ratio of distributions
$\pi_{\bf c}/\bpi_{\bf c}$, described in appendix \ref{appendixA}. Thanks to this factorization, we have been able to express the ${\cal D}_{p}$'s
as linear combinations among the relative entropies of the the marginal clusters distributions, as stated in \eqref{DpDef1}. In order to achieve our
main results a twofold effort has then been needed. Firstly, we have discussed the derivation of formulae \eqref{kPointPiDef1}, that express the marginal
probability distribution of a generic cluster in terms of its correlation functions. These equations allowed us to build an explicit representation
of the cluster relative entropies \eqref{CREDef1} in terms of correlation functions. Then, we have computed the coefficients $\a_{k}$ that weight
the contribution of the $k$-dimensional clusters. Once that these two intermediate results have been achieved, it has been possible to write the
multiparticle correlation expansion of the relative entropy directly from equation \eqref{DpDef1}. Some general properties of the result of this
construction, together with some explicit examples, have been discussed in section \ref{GeneralProperties}.

As a first application of this result, we have shown that the multiparticle correlation expansion of the excess entropy can be obtained from the
${\cal D}_{p}$ functions, by performing a proper choice of the reference distribution.  The link between relative entropy and excess entropy has been
established by means of general arguments in equation \eqref{ExcessEntropyExpDef2}. Following this relationship, we have built the ${\cal D}_{p}^{\sst{(0)}}$
functions, that express the multiparticle correlation expansion of the relative entropy between a generic homogenous system and a reference system realized
as the product of the one-point marginals of the former one. The analysis presented in section \ref{EntropyCE} evidenced that the multiparticle correlation
expansion of the excess entropy based on the ${\cal D}^{\sst{(0)}}_{p}$ functions produces results equivalent to the ones derived in
\cite{springerlink:10.1023/A:1004520432275}. In particular, we have shown  that the expression of ${\cal D}_{\sst{2}}^{\sst{(0)}}$ reported in formula
\eqref{D02pointdef1} contains the complete excess entropy $S_{\sst{2}}$ plus the two-body singular contributions at the higher orders in $\rho$
(equations \eqref{D02pointdef2} provide an explicit check of this statement up to $\rho^{\sst{4}}$). Analogous results holds at the level $p=3$, as stated
in \eqref{D03pointdef2}, and a further check has performed also for $p=4$. These results denote that ${\cal D}^{\sst{(0)}}_{p}$ can be interpreted as
a partial resummation of the series \eqref{ExcessEntropyExpDef3} limited to the contribution of clusters of maximum dimension equal to $p$.

%%%%%%%%%%%%%%%%%%%%%%%%%%%%%%%%%%%%%%%%%%%%%%%%%%%%%%%%%%%%%%%%%%%%%%%%%%%%

\appendix

\section{Derivation of the cluster decomposition formula for ${\cal D}_{p}$}
\label{appendixA}

The cluster expansion of the relative entropy given by equation \eqref{DpDef1} can be derived starting from an exact factorization
of the ratio of probability distributions $R_{\bf c}= \pi_{\bf c}/\bpi_{\bf c}$. This procedure is completely analogous to the
one that leads to the truncated cumulant expansion of the configurational entropy (see, for instance, \cite{Nettleton1958,Falk1971,
Sanchez1984334,Pelizzola2005}). The idea at the base of this approach is to express a generic function so that each term can be
ascribed to a distinct cluster of lattice sites. We start by defining the following representation for the logarithm of $R_{\bf c}$:
\be\label{FactorizationDef1}
\ln R_{c_{i_{1}}\cdots c_{i_{N}}} = \sum_{k=1}^{N} {\cal F}_{k}
\ee
where the index $k$ runs over the dimension of the clusters and the ${\cal F}_{k}$ terms are built in terms of the marginal ratios
$R_{c_{i_{1}}\cdots c_{i_{k}}}= \pi_{c_{i_{1}}\cdots c_{i_{k}}}/\bpi_{c_{i_{1}}\cdots c_{i_{k}}}$. The explicit form of the ${\cal F}_{k}$
terms reads:
\bea\label{FactorizationDef2}
&&{\cal F}_{\sst{1}} = \sum_{i=1}^{N} \ln R_{c_{i}}\nonumber \\
&&{\cal F}_{\sst{2}} = \sum_{i_{1}<i_{2}} \ln \left[\frac{R_{c_{i_{1}}c_{i_{2}}}}
{R_{c_{i_{1}}}R_{c_{i_{2}}}} \right] \nonumber \\
&&{\cal F}_{\sst{3}} = \sum_{i_{1}<i_{2}<i_{3}} \ln \left[\frac{R_{c_{i_{1}}c_{i_{2}}c_{i_{3}}}}
{R_{c_{i_{1}}c_{i_{2}}}R_{c_{i_{1}}c_{i_{3}}}R_{c_{i_{2}}c_{i_{3}}}}
R_{c_{i_{1}}}R_{c_{i_{2}}}R_{c_{i_{3}}} \right] \\
&& \dots \nonumber \\
&&{\cal F}_{k} = \sum_{i_{1}<\cdots<i_{k}} \ln \left[ \prod_{l=1}^{k} \prod_{j_{1}<\cdots<j_{l}}
\left(R_{j_{1}\dots j_{l}}\right)^{k-l} \right]
\nonumber
\eea
Where the last line of \eqref{FactorizationDef2} provides the explicit expression of the generic ${\cal F}_{k}$ term and is built by generalizing
the relations written for low values of $k$. The correctness of this formula can be checked by verifying that equation \eqref{FactorizationDef1}
is satisfied when ${\cal F}_{k}$ is plugged in its r.h.s. We see that, for each value of $k$, the sum over $i_{1}<\cdots<i_{k}$ selects all the
distinct $k$-dimensional clusters inside the $N$ lattice variables and the product over $j_{1}<\cdots<j_{l}$ runs on all the $l$-dimensional subsets
of a given cluster. The argument of the logarithm is written in terms of a combinations of marginal ratios, up to the  $k$-dimensional one, which in
turn represents the connected contribution of the selected cluster to $\ln R_{\bf c}$.

Equation \eqref{FactorizationDef1} describes an algebraic identity since the structure of the ${\cal F}_{k}$ terms guarantees that all
the contributions due to clusters of dimension lower than $N$ exactly cancel in the sum. At the same time it also represents the starting
point for a cluster expansion series. Indeed, we can limit the sum in its r.h.s. at the level of $p$-dimensional clusters and plug the
truncated expansion into the defining equation of the relative entropy \eqref{RelativeEntropyDef1}. In this way we obtain the approximated
expression of RE limited to the contribution of $p$-dimensional cluster, denoted as ${\cal D}_{p}$ in text. It reads:
\bea\label{FactorizationDef3}
&&{\cal D}_{p} = \sum_{\bf c} \pi_{\bf c} \left(\sum_{k=1}^{p} {\cal F}_{k} \right) = \nonumber \\
&& =  \sum_{i=1}^{N}\sum_{c_{i}}\pi_{c_{i}}\ln R_{c_{i}} +
\sum_{i_{1}<i_{2}}\sum_{c_{i_{1}},c_{i_{2}}}\pi_{c_{i_{1}}c_{i_{2}}}\ln \left[\frac{R_{c_{i_{1}}c_{i_{2}}}}
{R_{c_{i_{1}}}R_{c_{i_{2}}}} \right] + \dots \\
&& + \sum_{i_{1}<\dots<i_{p}}\sum_{c_{i_{1}},\dots,c_{i_{p}}}\pi_{c_{i_{1}}\cdots c_{i_{p}}} \ln
\left[ \prod_{l=1}^{p} \prod_{j_{1}<\cdots<j_{l}} \left(R_{j_{1}\dots j_{l}}\right)^{p-l} \right]
\nonumber
\eea
where the r.h.s has been written in terms of the marginals of $\pi_{\bf c}$. The final step needed to express ${\cal D}_{p}$ in the form
of equation \eqref{DpDef1} is realized by rewriting the logarithms of the connected ratios as the algebraic sum of the logarithm of each
factor, that is:
\be\label{FactorizationDef4}
\ln \left[ \prod_{l=1}^{k} \prod_{j_{1}<\cdots<j_{l}}
\left(R_{j_{1}\dots j_{l}}\right)^{k-l} \right] = \sum_{l=1}^{k} (-)^{k-l}\sum_{j_{1}<\cdots<j_{l}}
\ln \left(R_{j_{1}\dots j_{l}}\right)
\ee
Using this relation allows us to rewrite equation \eqref{FactorizationDef3} as a linear combination of the cluster relative entropies
${\cal C}_{k}$ associated to all the clusters of order ranging from 1 to $p$. Moreover, due to this rearrangement of the terms in
\eqref{FactorizationDef3}, each cluster is counted with a non trivial multiplicity since it can appear as subset of many higher dimensional
ones. Consequently, this analysis justifies formula \eqref{DpDef1} as the definition of the ${\cal D}_{p}$ functions and explains that
the $\alpha_{k}$ coefficients represent the algebraic multiplicity of the $k$-dimensional clusters in the sum \eqref{FactorizationDef3}.
These quantities are computed in section \ref{alphaCoefficients} by means of a combinatorial analysis.

%%%%%%%%%%%%%%%%%%%%%%%%%%%%%%%%%%%%%%%%%%%%%%%%%%%%%%%%%%%%%%%%%%%%%%%%%%%%

%%%%%%%%%%%%%%%%%%%%%%%%%%%%%%%%%%%%%%%%%%%%%%%%%%%%%%%%%%%%%%%%%%%%%%%%%%%%

\end{document}